# Estimating the density of resident coastal fish using underwater cameras: accounting for individual detectability

Guillermo Follana-Berná*[1,2,3], Miquel Palmer[3], Andrea Campos-Candela[3,4], Pablo Arechavala-Lopez[3], Carlos Diaz-Gil[1,2,3], Josep Alós[3], Ignacio A. Catalan[3], Salvador Balle[3], Josep Coll[5], Gabriel Morey[5], Francisco Verger[5], Amalia Grau[1,2]

1 Laboratori d'Investigacions Marines i Aqüicultura, LIMIA (Balearic Government), 07157, Port d'Andratx, Illes Balears, Spain

2 Instituto de Investigaciones Agroambientales y de Economía del Agua, INAGEA (INIA_Govern Balear-UIB), 07122, Palma, Illes Balears, Spain

3 Fish Ecology Lab, Instituto Mediterráneo de Estudios Avanzados, IMEDEA (CSIC-UIB), 07190, Esporles, Illes Balears, Spain.

4 Department of Marine Sciences and Applied Biology, University of Alicante, 03080, Alicante, Spain

5 Tragsatec, 07009, Palma, Illes Balears, Spain.

* Corresponding author: gfollana@dgpesca.caib.es

**Running page head:** Individual detectability and density with cameras.

# ABSTRACT

Technological advances in underwater video recording are opening novel opportunities for monitoring wild fish. However, extracting data from videos is often challenging. Nevertheless, it has been recently demonstrated that accurate and precise estimates of density for animals (whose normal activities are restricted to a bounded area or home range) can be obtained from counts averaged across a relatively low number of video frames. The method, however, requires that individual detectability ($P_{ID}$, the probability of detecting a given animal provided that it is actually within the area surveyed by a camera) has to be known. Here we propose a Bayesian implementation for estimating $P_{ID}$ after combining counts from cameras with counts from any reference method. The proposed framework was demonstrated using *Serranus scriba* as a case-study, a widely distributed and resident coastal fish. Density and $P_{ID}$ were calculated after combining fish counts from unbaited remote underwater video (RUV) and underwater visual censuses (UVC) as reference method. The relevance of the proposed framework is that after estimating $P_{ID}$, fish density can be estimated accurately and precisely at the UVC scale (or at the scale of the preferred reference method) using RUV only. This key statement has been extensively demonstrated using computer simulations yielded by real empirical data. Finally, we provide a simulation tool-kit for comparing the expected precision attainable for different sampling effort and for species with different levels of $P_{ID}$. Overall, the proposed method may contribute to substantially enlarge the spatio-temporal scope of density monitoring programs for many resident fish.

# 1. INTRODUCTION

One of the fundamental challenges in marine ecology and fisheries science is to describe the current state of fish populations in terms of abundance, which is imperative for understanding population dynamics (Hilborn & Walters 1992, Agnew et al. 2013). However, reliable abundance data at relevant spatio-temporal scales is rarely available in marine systems. Data scarcity is especially severe in the cases of recreational and artisanal fisheries targeting resident coastal fish, for which a science based, sustainable management uses tend to be unfeasible (Pita et al. 2018). In addition, even in well-monitored fisheries, most of the data comes from catches. However, the usefulness of fisheries dependent data (e.g., captures per unit effort; CPUE) as a proxy of abundance is under debate (Pauly et al. 2013). For example, catch-related data may be biasing against the less-vulnerable fish species or the less-vulnerable fraction within a given species (Alós et al. 2014, Alós et al. 2015, Alós et al. 2018). Further, both fish and fishermen behavior may induce a lack of proportionality between CPUE and fish density, producing hyperdepletion or hyperstability (Lennox et al. 2017).

The existence of bias when estimating abundance is widely recognized for either fishery-dependent or fishery-independent methods. Among fishery-independent methods (Murphy & Jenkins 2010, Mallet & Pelletier 2014, Przeslawski et al. 2018), scuba-diver's counting fish along standardized transects (i.e., underwater visual censuses or UVC) or point counts have been so widely used for monitoring many coastal fish species (Murphy & Jenkins 2010) that they currently represent standard, reference methods. UVC is a non-destructive method with many advantages but it suffers from some well-known drawbacks too. Biases may occur due to different behavioural responsiveness of fish species to the presence of divers (Lindfield et al. 2014), the observer-related effects (Dickens et al. 2011) or to within-transect variability (Kulbicki et al. 2010). Most of these problems are currently dealt with through using highly standardised protocols because biases are assumed to be consistent for a given species, thus allowing between-study comparisons (Ackerman & Bellwood 2000, Ward-Paige et al. 2010). Nevertheless, one of the main concern against UVC is the large investment in time and personal effort needed. The immediate outcome is that sample size tend to be small both at the temporal and at the spatial scales (Thompson & Mapstone 1997). Certainly, reduced sample sizes does not necessarily introduce bias but does imply worse precision and wider confidence intervals.

On the contrary, the use of underwater cameras or action underwater cameras is experiencing an increased interest because they supposedly allow long-term, high-frequency monitoring of fish and marine environments (Assis et al. 2013, Aguzzi et al. 2015, Easton et al. 2015), with the add value of avoiding any diver-related bias. The main advances that promote the use of cameras are miniaturization, drop in price, shock proofing, water proofing in a wide depth range, long-life batteries, high-capacity memory cards and high-definition images (Struthers et al. 2015). Additional potential applications of remote underwater video cameras (RUV) are, for example, its usefulness in marine reserves or for monitoring endangered species (Murphy & Jenkins 2010). In addition, RUV can cover a broader spatial and temporal scales (Pelletier et al. 2012, Assis et al. 2013), and potential biases due to the physical environment or to behavioural and life history characteristics might be also reduced.

However, RUV recording has also some disadvantages as a limited field of vision, the need for good visibility and the cost related with image processing (Pita et al. 2014, Struthers et al. 2015). In most cases, bait is used for attracting fish around the camera (Whitmarsh et al. 2017). Nevertheless, several problems precludes any reliable estimation of absolute density using baited cameras (Whitmarsh et al. 2017). Namely, the unknown dynamics of the bait odour plume, how such dynamics is affecting attractiveness (Dunlop et al. 2015), how the fish already attracted by the bait are themselves a visual cue for other fish, and the plausible existence of species-specific responses and internal state dependence of the individual. Instead, several relative abundance metrics have been proposed. Two of the most popular metrics are MaxN and MeanCount (Stobart et al. 2007, Schobernd et al. 2014, Campbell et al. 2015, Stobart et al. 2015). This type of metrics have been developed because most ecologists and managers are concerned with counting the same individual more than once (Ward-Paige et al. 2010, Assis et al. 2013, Campbell et al. 2015).

Alongside the technological opportunities offered by camera-based wildlife assessment, it was recently demonstrated that fish density (number of animals per area unit) can be properly estimated from fish counts across relatively few video frames obtained whit an unbaited camera (Campos-Candela et al. 2018). One of the key assumptions of the method proposed by these authors is that animal density do not change at the spatial and temporal scale of a given RUV sample. Fortunately, this assumption meets for many resident coastal fish that remain most of the time within a given area of activity, or home range (HR), which tends to be orders of magnitude smaller than the extent of suitable habitat (March et al. 2010, Villegas-Rios et al. 2014, Alós et al. 2016). For those HR behaving fish, no emigration, no immigration, no changes in the HR location, no birth and no death can be safely assumed at the spatial and temporal scales commonly used for sampling abundance.

Moreover, an additional assumption of the previously mentioned method is that Individual Detectability ($P_{ID}$) has to be known (Campos-Candela et al. 2018). We define $P_{ID}$ here as the probability of counting a given fish that is actually within the area sampled by a camera. Conspicuous species are more easily detected than cryptic species (Boulinier et al. 1998) and, even the same individual will be more detectable at a bare habitat than at complex habitats where fish may be hidden behind rocks or within seagrass patches. Moreover, the existence of a positive relationship between detectability and density can artificially inflate between surveys heterogeneity (Dorazio & Royle 2005). Hence, the development of a method accounting for imperfect detection is indispensable for making available unbiased inferences on fish density when using RUVs (Bacheler et al. 2017). Note that $P_{ID}$ should not be confused with the concept of detectability used in ecology and conservation biology to infer if a given species inhabits a given site even when it has not been detected there (Boulinier et al. 1998, Bayley & Peterson 2001).

The aim of this paper is to propose and demonstrate a method for the concurrent estimation of density and $P_{ID}$ for species that (i) display HR behaviour, (ii) for which UVC represents a reliable reference method and (iii) for which unbaited RUV is a feasible alternative (Campos-Candela et al. 2018). The relevance of the method proposed here relies in the fact that after estimating $P_{ID}$, RUV can be used alone for producing low cost abundance estimates at the scale of the reference method (here UVC), which may entail a paradigm shift for monitoring fish density at large spatial and temporal scales. For demonstrating the method, a HR behaving Serranid, *Serranus scriba*, was selected as case study. Accuracy and precision of the method was evaluated using computer simulation experiments (i.e., moving virtual fish according with a reliable movement model (March et al. 2010, Alós et al. 2016). Simulations settings were realistic in terms of fish movement, feasible fish densities and diverse $P_{ID}$. Finally, we provided a simulation tool for exploring the precision attainable with different sampling effort and with different $P_{ID}$. This tool can guide for optimizing the sampling strategies in the field.

## 2. MATERIALS & METHODS

## 2.1 Theoretical framework

The logical rationale behind the method proposed here is that fish density is the same when sampling the same site with RUV and with the reference method. Therefore, after proper adjusting for RUV detectability, RUV fish counts may provide either absolute density estimates when individual detectability of the reference method is perfect (i.e., all fish are detected) or density estimates at the scale of the reference method otherwise.

Under the assumptions that (i) animals move independently from one another, and (ii) the density is stationary at the sampling temporal and spatial scale (Campos-Candela et al. 2018), the counts obtained with the reference method (NREF) are expected to be Poisson-distributed:

$$N_{REF} \sim Poisson(\lambda_{REF}) \hspace{4cm} \text{Eq. 1}$$

where $\lambda_{REF}$ is given by:

$$\lambda_{REF} = P_{ID.REF} D Z_{REF} = D_{REF} Z_{REF} \hspace{3cm} \text{Eq. 2}$$

Where $P_{ID.REF}$ is the individual detectability of the reference method, D is the (true) animal density (animals per area unit) and $Z_{REF}$ is the area sampled with the reference method. Therefore, $D_{REF}$ is the (relative) density at the scale of reference method after accounting for $P_{ID.REF}$.

Similarly, the counts obtained with RUV are expected to be Poisson-distributed:

$$N_{RUV} \sim Poisson(\lambda_{RUV}) \hspace{4cm} \text{Eq. 3}$$

where $\lambda_{RUV}$ is given by:

$$\lambda_{RUV} = P_{ID} D Z_{RUV} = \frac{P_{ID.RUV}}{P_{ID.REF}} D_{REF} Z_{RUV} = P_{ID} D_{REF} Z_{RUV} \hspace{1.5cm} \text{Eq. 4}$$

Where $P_{ID.RUV}$ is the RUV detectability relative to the $P_{ID.REF}$ of the reference method. Therefore, after estimating $P_{ID}$, fish density at the scale of the reference method can be estimated using RUV only. Note that if $P_{ID.REF} = 1$ (all fish are detected by the reference method), then RUV counts are providing an unbiased estimate of the absolute density.

## 2.2 Specificities of the study case

To demonstrate the proposed framework data were collected from 5 sites along the SW coast of Mallorca (Fig 1), where 51 UVCs were completed and 13 RUVs were deployed (Supplementary material 1). UVCs were conducted by three different scuba divers, between 5 and 25 m depth and between 9:00 and 12:00 GTM, covering an area of 250 m2 (50 m long and 5 m wide) during 25 minutes. Each diver completed as maximum four transects per day at different sites (Supplementary material 1). Nine UVCs were completed per site, excepting one site where 15 UVCs were completed. Transects were located at least 20 m apart each other to minimise spatial autocorrelation (Ordines et al. 2005). The number of *S. scriba* at each UVC was recorded. Note that UVCs were specifically optimized for counting *S. scriba*.

RUV device was a stainless steel structure equipped with two stereoscopic action cameras (Supplementary material 2). RUVs were not baited. RUV structures were deployed around the UVCs (Supplementary material 3). They were deployed during 4 hours (9:00 and 14:00 GTM) but the maximum duration of the batteries was approximately 3 hours only. Three RUVs were deployed per site but two videos were discarded. Failures were due to battery problems, poor visibility or bad deployment. Excepting in those failures cases, RUVs were deployed at approximately 25 meters apart of the corresponding UVC. More detail are supplied at supplementary material 1 and 3.

Concerning video analysis, the five first minutes were excluded in order to avoid any disturbance related with the deployment itself. All individuals of *S. scriba* were counted in each frame every 150 seconds (or every 9000 frames). Preliminary analyses showed that this frequency minimizes temporal autocorrelation between frames. The average number of frames counted per video was 73. The distance of any given fish to the RUV was estimated using a Matlab Calibration Toolbox (Díaz-Gil et al. 2017). Fish were only counted when they were at less than 2.5 m from the RUV. Preliminary tests suggest that at this distance, detectability did not depend on fish size. Provided that the view angle of the camera was 127°, the area surveyed was 6.93 m2.

2.3 Modelling fish density and detectability.

The fish counts for a given site, diver and day (N.UVCsite,diver,day.UVC) was expected to be Poisson distributed around a mean value λ.UVCsite,diver,day.UVC. Site-related effect ($\mu.UVC_{site}$) was considered as a fixed factor, while diver-related effect ($\delta_{diver}$) and day.UVC-related effect ($\delta_{day.UVC}$) were considered random factors.

$$N.UVC_{site,diver,day.UVC} \sim Poison\big(\lambda.UVC_{site,diver,day.UVC}\big) \qquad \text{Eq. 5}$$

$$\log\big(\lambda.UVC_{site,diver,day.UVC}\big) = \mu.UVC_{site} + \delta_{diver} + \delta_{day.UVC} \qquad \text{Eq. 6}$$

$$e^{\mu.UVC_{site}} = D_{REF}Z_{UVC} \qquad \text{Eq. 7}$$

$$\delta_{diver} \sim Normal(0, \sigma_{diver}) \qquad \text{Eq. 8}$$

$$\delta_{day.UVC} \sim Normal\big(0, \sigma_{day.UVC}\big) \qquad \text{Eq. 9}$$

where $Z_{UVC}$ is the area of the UVC and $D_{REF}$ is the density at the scale of the reference method in the sampled site, in this case UVC.

Concerning RUVs, the fish count for frame i at a given site, camera and day ($N.RUV_{i,site,cam,day}$) is assumed to be Poisson distributed around a mean value $\lambda.RUV_{site,cam,day}$. Site-related effect ($\mu.RUV_{site}$) was considered as a fixed factor, while camera-related effect ($\delta_{cam}$) and day.cam-related effect ($\delta_{day.cam}$) were considered random factors.

$$N.RUV_{i,site,cam,day.cam} \sim Poisson\left(\lambda.RUV_{site,cam,day.cam}\right) \qquad \text{Eq. 10}$$

$$\log\left(\lambda.RUV_{site,cam,day.cam}\right) = \mu.RUV_{site} + \delta_{cam} + \delta_{day.cam} \qquad \text{Eq. 11}$$

$$e^{\mu.RUV_{site}} = P_{ID} D_{REF} Z_{RUV} \qquad \text{Eq. 12}$$

$$\delta_{cam} \sim Normal(0, \sigma_{cam}) \qquad \text{Eq. 13}$$

$$\delta_{day.cam} \sim Normal\left(0, \sigma_{day.cam}\right) \qquad \text{Eq. 14}$$

where $Z_{RUV}$ is the detection area of the camera, $P_{ID}$ is the probability of detection and $D_{REF}$ is the density at the scale of the reference method in the sampled site.

The parameters of the model (eqs. 5 to 14) given the observed data were fitted using a Bayesian approach. Samples from the joint posterior distribution of parameters (specifically, from D and $P_{ID}$) given the data (fish count from UVC and RUV), were obtained using JAGS (http://mcmc-jags.sourceforge.net/ accessed 20 Dec 2018) (Plummer 2015) and the r2jags library (Su & Yajima 2015) of the R package (R Core Team 2017 at http://www.r-project.org/ accessed 20 Dec 2018). Non-informative priors were assumed according to symmetric probability distributions and previously published data: a uniform prior between 0 and 1 fish m$^{-2}$ for D (García-Charton and Pérez-Ruzafa 2001, Arechavala-López et al. 2008, Deudero et al. 2008), a uniform prior between 0 and 1 for $P_{ID}$, and uniform prior between 0 and 10 for the standard deviation of all random effects. Three Monte Carlo Markov Chains (MCMC) were run. We drew 30000 posterior samples. The first 15000 iterations were discarded and only one out 10 of the remaining iterations were kept in order to prevent autocorrelation. MCMC convergence was assessed by visual inspection and evaluated using the Gelman-Rubin Statistic (Plummer et al. 2006). The detailed model design, the R script and a user-friendly interface to derive the parameters for any set of data can be found at the free Shiny application website: https://fishecology.shinyapps.io/uvccam/ (accessed 20 Dec 2018).

2.4 Computer simulation experiments

The relevance of the framework proposed here is that, after estimating $P_{ID}$ with the method described in section 2.3, fish density at new sites can be accurately and precisely estimated using RUV only. This statement has been demonstrated using four *sets* of simulations consisting in moving fish in a virtual scenario where *virtual cameras* were deployed.

For a given simulation *set*, 100 *replicates* of 10 *virtual cameras* each were considered. Fish density was estimated for each *replicate* using the model

$$N.RUV_{i,cam} \sim Poisson(\lambda.RUV_{cam}) \qquad \text{Eq. 15}$$

$$\log(\lambda.RUV_{cam}) = \mu.RUV + \delta_{cam} \qquad \text{Eq. 16}$$

$$e^{\mu.RUV} = P_{ID}D_{REF}Z_{RUV} \qquad \text{Eq. 17}$$

$$\delta_{cam} \sim Normal(0, \sigma_{cam}) \qquad \text{Eq. 18}$$

Where $P_{ID}$ was not estimated but assumed to be known.

We generated movement trajectories of fish displaying home range using the model used in (Palmer et al. 2011, Alós et al. 2016, Campos-Candela et al. 2018):

$$\vec{r}_{n+1} = \vec{r}_c + e^{-k\Delta t}(\vec{r}_n - \vec{r}_c) + \vec{R}_n \qquad \text{Eq. 19}$$

Where $\vec{r}_{n+1}$ denotes the position at discrete time $t_{n+1} = (n + 1)\Delta t$, $\vec{r}_n$ denotes the current position (Cartesian coordinates) of the fish at time $t_n = n\Delta t$. $\vec{r}_c$ is the position of the centre of the HR, k is a central harmonic constant force attracting the animal towards $\vec{r}_c$, and $\vec{R}_n$ is a stochastic term, normally distributed with zero mean and standard deviation ($\sigma$) in each dimension approximated by (Palmer et al. 2011):

$$\sigma = \sqrt{\frac{\varepsilon\left(1 - e^{-2k\Delta t}\right)}{2k}}$$
Eq. 20

The values for k (0.258 s-1) and ε (631.05 m2 s-1) used for moving fish were those estimated for *S. scriba* by acoustic tracking (March et al. 2010, Campos-Candela et al. 2018). The time step Δt at which the position of all fish were updated was set to 1 second.

Each one of the 10 *virtual cameras* of a given *replicate* was set at the centre of a squared virtual scenario with side defined as twice the radius of the HR (rHR). In the case of *S. scriba*, rHR was estimated in 85.6 m using acoustic tracking (March et al. 2010). The rationale of using such a side length is that an animal having its centre of HR outside the scenario considered (and, thus, not included in the simulation) has a negligible probability of being detected by a *virtual camera*. The number of animals moved within such a scenario is given by side2D, where D is the fish density actually estimated for *S. scriba* (see results). The centres of HR of the virtual fish were randomly distributed within the virtual scenario.

A given *virtual camera* was simulated to be deployed for 10 hours; thus, the number of fish movements tracked for any given fish was 360000. However, the number of fish that were within the area surveyed by the camera was counted at one time step every 150 seconds. The number of fish counted was randomly sampled from a binomial distribution of the actual number of fish with probability defined by $P_{ID}$, which is assumed to be known.

The virtual simulations are designed for demonstrating the outcomes of accounting for $P_{ID}$. The four simulations *sets* differed either in the value of $P_{ID}$ considered for counting a fish ($P_{ID,sim}$) and in the way $P_{ID}$ is modelled ($P_{ID,model}$).

First *set*: $P_{ID,sim}$ was set to 1 (all fish actually within the area surveyed by the camera are counted) and $P_{ID,model}$ was rightly assumed to be 1.

Second *set*: $P_{ID,sim}$ was set to the value estimated here ($P_{ID} = 0.65$; see results section) but $P_{ID,model}$ was wrongly assumed to be 1 (i.e. ignoring the potential bias related with imperfect detectability).

Third *set*: $P_{ID,sim}$ was set to 0.65 and $P_{ID,model}$ was rightly assumed to be 0.65. This *set* emulates the case in which $P_{ID}$ has been previously estimated in a pilot experiment using the protocol described in sections 2.2 and 2.3. After it, fish density are estimated at new sites using cameras only.

Fourth *set*: $P_{ID,sim}$ was set to 0.65 but the uncertainty when estimating *PID* was explicitly accounted for:

$$\text{logit}(P_{ID}) \sim Normal(mean, sd) \qquad\qquad \text{Eq. 21}$$

Where *mean* and *sd* are the mean and standard deviation of the logit-transformed posterior values of the $P_{ID}$ estimated here ($P_{ID,model} = 0.65$, with a 95% Bayesian Credibility Interval between 0.34 and 0.95).

Accuracy of the estimated D at each simulation *set* was assessed by the scaled root mean squared error (SRMSE; Walther and Moore 2005).

$$SRMSE = \frac{1}{Dr} \sqrt{\frac{1}{n} \sum_{j=1}^{n} \left( Ds_j - Dr \right)^2} \qquad \text{Eq. 22}$$

where $Dsj$ is the estimated value of density in the $j$ simulation and $Dr$ was the true value. The inter-quantile (2.5% to 97.5%) range of the posterior Bayesian Credibility Intervals was used for assessing precision of D.

## 3. RESULTS

### 3.1 Empirical data

Concerning the field experiments, after combining fish counts from UVC and RUV, the estimated median fish density across five sites ranged between 0.016 and 0.017 fish $m^{-2}$. Fish density appears to be the same across sites, provided that credibility intervals largely overlap (Fig. 2). The estimated values accounting for the different sources of variability are detailed at Table 1. Concerning $P_{ID}$, the 95% credibility interval ranged between 0.34 and 0.95, being 0.65 its median value. Thus, ignoring detectability may be a relevant concern for the specie studied and with the current RUV setting.

### 3.2 Simulation data

The relevance of taking into account $P_{ID}$ when estimating fish density using RUV data only was clearly demonstrated with the results obtained from the four simulations *sets* because precision and accuracy of the estimated densities can be compared with the true (simulated) fish density, which was 0.016 fish $m^{-2}$ (section 3.1).

The first simulation *set* emulates the case in which any fish within the area surveyed by the camera in a given frame is detected and counted ($P_{ID.sim} = 1$ and $P_{ID.model} = 1$). As expected, the estimated fish densities in that case were very accurate after a relatively small sample effort (Fig. 3). After 73 frames, (i.e., the number of frames used in the fieldwork) the average density (from 100 *replicates*) was 0.017 (interquartile range between 0.015 and 0.018). Moreover, precision was excellent too. For example, in a given *replicate* (a *set* of 10 cameras), the 95% confidence interval was between 0.013 and 0.021. Moreover, 95% CI included the true value (0.016 ind m$^{-2}$) in all 100 *replicates*.

The second simulation *set* emulates the case in which only an average of 65% of the fish that are actually within the area surveyed by the camera in a given frame are actually counted ($P_{ID,sim} = 0.65$) but this imperfect detection is ignored when estimating fish density from those counts ($P_{ID,model} = 1$). As expected fish densities was underestimated and the size of the bias was around 35% (i.e., $1 - P_{ID,sim}$). Note that the same pattern was obtained irrespective of the sampling same effort. After 73 frames, (i.e., the number of frames used in the fieldwork) the average density (from 100 *replicates*) was 0.011 (interquartile range between 0.010 and 0.012). In this case, the precision was excellent but the estimates were biased even after increasing the sampling effort in term of the number of frames included in the analyses. For example, in a given *replicate* (a *set* of 10 cameras), the 95% confidence interval was between 0.008 and 0.015. Moreover, 95% CI included the true value (0.016 ind m$^{-2}$) in only the 24 % of the 100 *replicates*.

The third simulation *set* emulates the case in which only an average of 65% of the fish that are actually within the area surveyed by the camera are counted ($P_{ID,sim} = 0.65$). However, and contrary with the second simulation *set*, in this case it is assumed that $P_{ID,model}$ has been previously estimated using protocol describe in section 2.2 and 2.3. Then this value of $P_{ID,model}$ was included as input when estimating fish density. In that case, the estimated fish density was no longer biased as in the simulation *set* 2 but accurate. After 73 frames, (i.e., the number of frames used in the fieldwork) the average density (from 100 *replicates*) was 0.017 (interquartile range between 0.015 and 0.018), very similar to the values obtained in the simulation *set* 1. Similarly, precision was excellent too. For example, in a given *replicate* (a *set* of 10 cameras), the 95% confidence interval was between 0.012 and 0.023. Moreover, 95% CI included the true value (0.016 ind m$^{-2}$) in all 100 *replicates*.

Finally, the fourth simulation *set* is very similar to the third simulation *set*. The unique difference in this *set* is that the uncertainty in $P_{ID,model}$ has been accounted for. In that case, the estimated fish density was accurate but certainly, the uncertainty in $P_{ID,model}$ is translated into worse precision estimates for fish density. After 73 frames, (i.e., the number of frames used in the fieldwork) the average density (from 100 *replicates*) was 0.017 (interquartile range between 0.015 and 0.018), very similar to those obtained in the simulation *set* 1 and 3. Simultaneously, precision was not as excellent as in the third *set* simulation; in this case, the precision was wider. For example, in a given *replicate* (a *set* of 10 cameras), the 95% confidence interval was between 0.010 and 0.047. Moreover, 95% CI indeed the true value (0.016 ind m$^{-2}$) in all 100 *replicates*.

Note that this is the more realistic simulation *set*. In fact, we strongly suggest completing a similar simulation exercise in order to assess the sampling effort in terms of cameras and deployment time needed for full fill a target at precision with the values of density and $P_{ID}$ estimated in a previous pilot study. The script used for the simulations and an easy-to-use Shiny app are included to facilitate these simulations (https://fishecology.shinyapps.io/uvccam/ accessed 20 Dec 2018).

For completeness, a sensitivity analysis was carried out with different detectabilities and different densities to inquire into how these variables affected the accurate and precise of RUV-only estimates of fish density. The sensitivity analysis shows that when the density of individuals in the environment and the $P_{ID}$ increase, the effort in number of frames needed to extract the real density using RUV decreases. In addition, from detectabilities close to 25% it is possible to obtain the density of animals with an effort in number of frames relatively low, even when the real density of the environment is low (0.01 ind m$^{-2}$) (Supplementary material 4).

## 4. DISCUSSION

We have successfully demonstrated a new framework that combines UVC and RUV for the concurrent estimation of fish density (individuals per area unit) and $P_{ID}$. The estimated fish density of *S. scriba* in the South coast of Mallorca (0.016 ind m$^{-2}$; 95% CI 0.011 - 0.027) is within the range reported using other methods, from other sites in Mallorca (Deudero et al. 2008) or in the Western Mediterranean (García-Charton and Pérez-Ruzafa 2001, Arechavala-López et al. 2008).

Nevertheless, the results extracted from our study-case are relatively irrelevant when compared with the opportunities that offers our method with the possibility of estimating $P_{ID}$. As mentioned above, the relevance of the proposed framework is that, after estimating $P_{ID}$, fish density can be estimated accurately and precisely at the scale of the preferred reference method using RUV only. Thus, provided that none of the commonly used reference methods for estimating density of resident coastal fish is precise enough, the possibility of an extensive use of properly calibrated RUV may contribute to substantially enlarge the spatio-temporal scope of density monitoring programs for many of resident coastal fish species.

Underwater visual census (UVC) is one of the most used methods by scientists, managers and stakeholders for estimating density of many species of coastal fish. The pros and cons of UVC have been comprehensively listed above, but here we emphasize that the time and effort cost precludes large sample sizes, thus densities are usually estimated with wide confidence intervals, thereby ecologically relevant processes may remain undetected. RUV cost is misperceived as similar or even large, especially when including video post processing (Pita et al. 2014, Murphy & Jenkins 2010). However, the cost per sample is by far more cost-effective. Moreover, recent advances in machine learning foresee cost reduction after developing applications for unsupervised recognition of fish from videos (Boom et al. 2014, Hsiao et al. 2014, Sun et al. 2018). The major problem of RUVs is that, in general, not all the fish that are actually within the area surveyed by the camera are detected (i.e., $P_{ID}$ tend to be less than 100%), thus fish density tend to be underestimated in relation to the preferred reference method.

Therefore, combining the advantages of UVC and RUV is certainly appealing. The need of combining data gathered using different methods have been repeatedly claimed. Specifically, simultaneous use of RUV and UVC have been extensively advised (Willis et al. 2000, Cappo et al. 2003, Stobart et al. 2007, Murphy & Jenkins 2010, Harvey et al. 2013, Shortis et al. 2013, Bacheler et al. 2017, Bosch et al. 2017). However, in most of those cases the results obtained using different methods are not combined but compared only (Cappo et al. 2004, Assis et al. 2013, Tessier et al. 2013, Pita et al. 2014, Mallet et al. 2014, Bacheler et al. 2017). Conversely, the framework proposed here offers the possibility for a genuine combination of fish counts using RUV with fish counts obtained with some other reference method to provide a more realistic view of the fish densities in coastal areas. Moreover, the Bayesian approach proposed here could be easily adapted to the specificities of any sampling strategy (e.g., including the confounding effects of covariables or different levels of random factors, such as between day or between-cameras effects), and our simulation tool-kit allows for extensive pre-sampling optimal settings (number of cameras and recording time) selection.

Nevertheless, we are not only proposing to combine RUVs with the preferred reference method for improving the precision of the estimated density at any new site or moment. Certainly, this may be an alternative in some cases but, instead, we propose to go one step further: Firstly, to concurrently conduct RUVs and UVCs in a single or in a few preliminary experiments for estimating $P_{ID}$ (according with the protocol detailed along sections 2.2 and 2.3). Secondly, after such a calibration exercise, only to deploy a large number of cameras for estimating density at several sites and moments for covering large spatial and temporal scales. According with the simulation experiments reported here, this might be a reliable alternative. Note that the need of inter-calibration exercises has been repetitively advised for comparing different methods of underwater camera surveys (Watson et al. 2005) or when suggesting that fish counts at a given UVC should be made by more than one scuba-diver at the same time (Bernard et al. 2013). However, the framework proposed here has a broader applicability. The extensive simulation experiments completed here suggest that density estimates may be accurate even when $P_{ID}$ is relatively small or even when it has been inaccurately estimated (Supplementary material 4). Indeed, in the latter case (Fig. 3), the sampling effort must be larger for attaining a target precision and the question of how optimally enlarge the sampling effort is an elusive topic. We strongly suggest completing a pilot experiment for identifying the levels of larger variability. For example, in the case study reported here, the large between-camera variance suggest the existence of heterogeneous habitats at the within-site level, thus advising to increase the number of cameras per site and perhaps to reduce the number of frames surveyed per camera. This could be achieved either by reducing the deployment time or by enlarging the time between two consecutive frames. The sensitivity analyses (Supplementary material 4) report the main patterns expected after changes in fish density and detectability, and estimate the expected cost (in terms of sampling effort) needed for achieving a target precision. Moreover, the R scripts and the Shiny app provided here may help for finding the optimal allocation of sampling effort for any given case.

Some specificities of the case study here deserves special attention. *S. scriba* moves relatively slowly within the area surveyed by the divers, thus, it is reasonable to assume that UVCs were synoptic in the sense that no fish enter or exit the area surveyed during the sampling period (Ward-Paige et al. 2010). Moreover, in our case, UVC were specifically designed for counting *S. scriba* only and special care was invested in searching within cavities and seagrass patches. Under these or similar specific circumstances, it may be feasible to assume that UVCs are detecting close to 100% of the individuals. In those specific cases, fish counts from RUV only and after calibrating for $P_{ID}$ will render absolute density (fish m$^{-2}$), as indicated in equation 4.

Note also that the confidence interval for $P_{ID}$ in the study case here is wide (from 0.34 to 0.95). Imprecise estimation of $P_{ID}$ translates into imprecise density estimates, thus, there are still place for many improvements. Increasing sampling effort when calibrating $P_{ID}$ would certainly increase $P_{ID}$ precision. This may be achieved by increasing the number of UVCs (or the area surveyed), by increasing the number of RUV, the deployment time, the number of frames analysed by each camera, or the area surveyed by the camera. In addition, some improvements concerning the RUV design may be advantageous. For example, when RUV looks parallel to the sea bottom (as those used in the experiments completed here), some fish may remain hidden behind a rock or a vegetation patch, which certainly increase uncertainty. Conversely, RUV looking down to the bottom may improve not only $P_{ID}$ but also its precision.

Moreover, a number of variables may affect $P_{ID}$ and should be accounted for. First of all, cryptic species should not be considered as optimal candidates for the framework reported here. Moreover, detectability of the very same individual may be habitat-dependent (complexity of the bottom structure). Seasonal differences in fish behaviour (i.e., fish may be more active and thus be more easily detected at warmer seasons), any other behaviour specificity or even fish size may affect detectability. Therefore, it is mandatory to take into account the behavioural attributes of the species studied prior to selecting a specific UVC and/or RUV design (Cheal & Thompson 1997, Samoilys & Carlos 2000, Ward-Paige et al. 2010). Provided that the area surveyed by the camera is usually small, it should be environmentally homogeneous too. However, from the UVC side, unaccounted sea bottom heterogeneity within a given transect is expected to increase the unaccounted variance in fish counts, which will translate to $P_{ID}$. Thus, many small but environmentally homogeneous UVCs would be preferred against a few large but heterogeneous UVCs (Murphy & Jenkins 2010). As mentioned above, UVCs should be as synoptic as possible which would ultimately depends on both, fish and diver speed (Ward-Paige et al. 2010, Pais & Cabral 2017). The design of RUV and UVC should be carefully selected to minimize all the potential confounding effects mentioned above. Moreover, habitat-specific $P_{ID}$ or similar dependencies can be empirically estimated after carefully designed calibration experiments.

Therefore, after solving those case-specific challenges, the methodological framework proposed here suggests that RUV surveys might be included in the basic toolkit in order to produce more reliable abundance estimates, thus enabling improvements in the management of coastal fish populations.

## Acknowledgements

GFB and CDG were supported by a PhD fellowship (FPI-INIA) from the National Institute for Agricultural and Food Research and Technology (INIA). ACC was supported by a Spanish FPU PhD fellowship (ref.FPU13/01440). PAL was supported by a Juan de la Cierva Incorporación postdoctoral grant (ref.IJCI-2015-25595). JA was supported by a Juan de la Cierva Incorporación postdoctoral grant (ref. IJCI-2016-27681). This work was funded by R+D project PHENOFISH (ref.CTM2015-69126-C2-1-R; MINECO) and is a contribution of the Joint Research Unit IMEDEA-LIMIA. This study was carried out with permission from the fisheries local administration, Government of the Balearic Islands. We also specially thank the researchers and students involved in the fieldwork.

| Sites | B.C.I. 2.5% | Median | B.C.I 97.5% | Rhat |
|---|---|---|---|---|
| Density [1] | 1.1 | 1.7 | 2.7 | 1.004 |
| Density [2] | 1.0 | 1.6 | 2.7 | 1.003 |
| Density [3] | 1.1 | 1.7 | 2.8 | 1.008 |
| Density [4] | 1.1 | 1.8 | 2.9 | 1.001 |
| Density [5] | 1.0 | 1.6 | 2.5 | 1.004 |
| $P_{ID}$ | 0.34 | 0.65 | 0.95 | 1.001 |
| sd.cam | 0.116 | 0.509 | 0.975 | 1.035 |
| sd.day.cam | 0.116 | 0.509 | 0.975 | 1.035 |
| sd.diver | 0.116 | 0.509 | 0.975 | 1.035 |
| sd.day.UVC | 0.026 | 0.105 | 0.383 | 1.001 |

Table 1. Estimated values of: Density (ind 100m$^{-2}$) of the five sites surveyed, Individual Detectability ($P_{ID}$) and variability of random effects (cam, day.cam, divers and day.UVC). B.I.C. is the Bayesian Credibility Interval. Rhat is the potential scale reduction factor and explain how the chains have converged to the equilibrium distribution. Approximate convergence is diagnosed when the upper limit is close to 1.

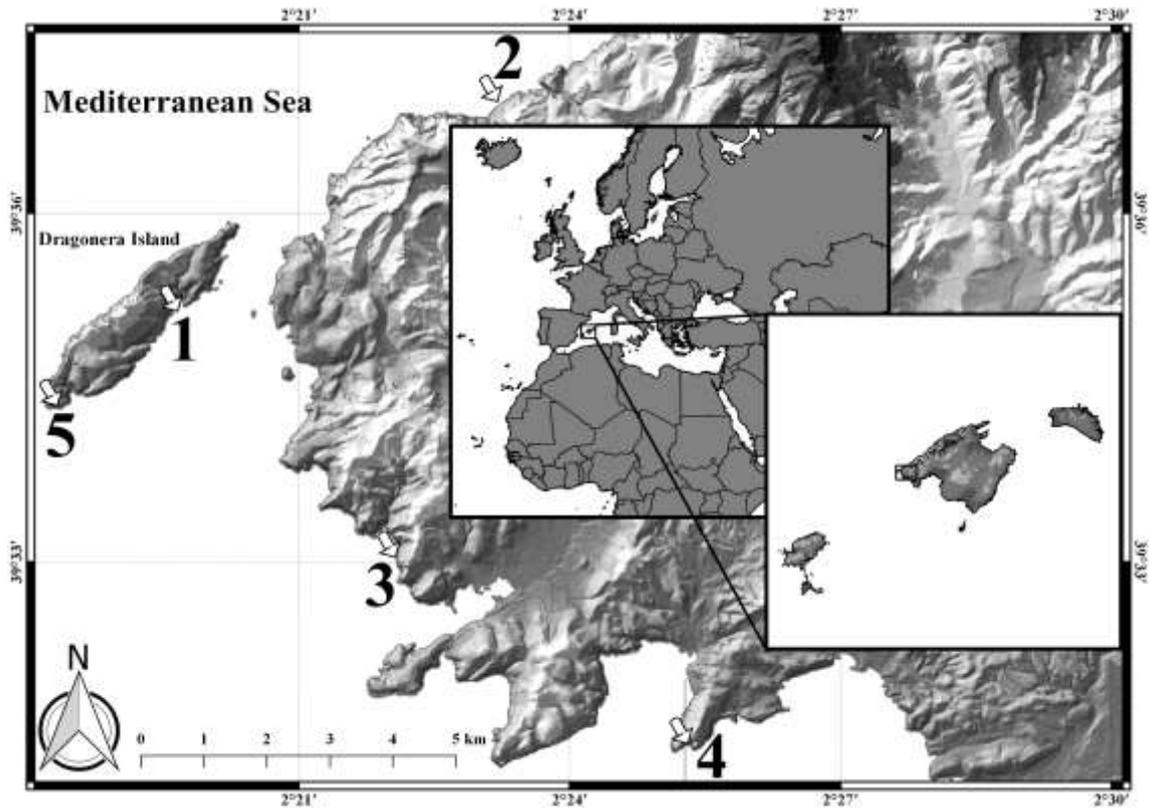

Fig 1. Map of the southwest part of the Mallorca Island indicating sampling sites. In supplementary material 1 you can see each site specifically. The numbers matches the number in Table 1 and Fig. 2 of results.

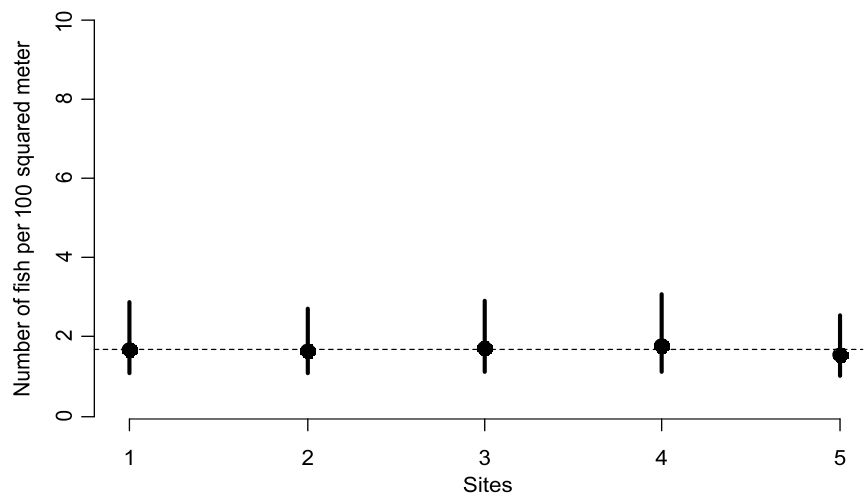

Fig 2. Estimated density combining UVC and RUV with its Bayesian credibility intervals (95%) at the five sites surveyed in the SW of Mallorca Island. Dot line is the mean value of the median (50%) density of the five sites.

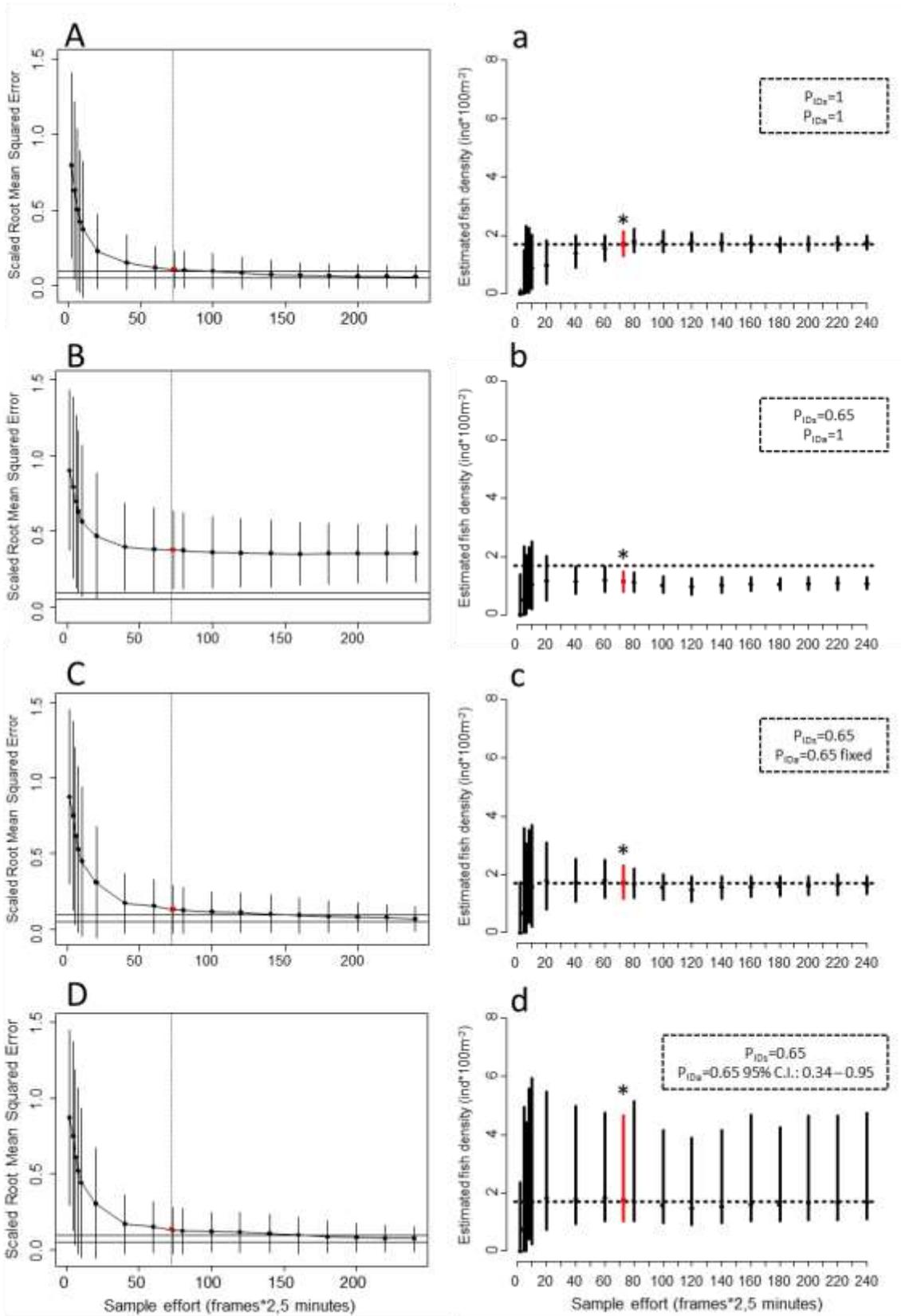

Fig 3. The plots with capital letters represent the Scaled Root Mean Squared Error (SRMSE) (Y-axis). SRMSE is the accuracy of density estimates. X-axis represent the increasing effort in number of frames (frames analysed by 10 cameras). The continuous horizontal lines correspond to threshold values of 10% and 5% of the SRMSE. The plots with lower case represent the precision for the estimated density with increasing sampling effort (number of frames analysed by 10 cameras). Mean values (black points) and 95% Bayesian credibility intervals (BCIs) of the medians from posterior distributions of the density estimates for 10 cameras. The dotted line correspond to the true density value obtained from empirical data combining UVC and RUV. The asterisks indicate the effort, in frames, used in the fieldwork. In the square of points is described the Individual detectability used in the simulation ($P_{ID,sim}$) and analysis ($P_{ID,model}$).

Supplementary material 1. The tables describe the number of RUVs and UVCs per days and sites when the surveys were done. In addition, the total number of RUVs and UVCs per site (horizontal), per day (vertical) are showed and the total number of RUVs and UVCs realized in the survey (lower right corner).

**RUV** **Day**

| Sites | 20/07/2016 | 22/07/2016 | 08/08/2016 | 12/08/2016 | 16/08/2016 | 17/08/2016 | 26/10/2016 | Total |
|---|---|---|---|---|---|---|---|---|
| 1 | 1 | 1 | | | | 1 | | 3 |
| 2 | 1 | | 1 | | | | | 2 |
| 3 | | 1 | | | 1 | | 1 | 3 |
| 4 | | 1 | | 1 | 1 | | | 3 |
| 5 | | | 1 | | | | 1 | 2 |
| **Total** | 2 | 3 | 2 | 1 | 2 | 1 | 2 | 13 |

**UVC** **Day**

| Sites | 08/07/2016 | 11/07/2016 | 12/07/2016 | 14/07/2016 | 15/07/2016 | 16/07/2016 | 18/07/2016 | Total |
|---|---|---|---|---|---|---|---|---|
| 1 | | | 3 | 3 | 3 | | | 9 |
| 2 | | | 3 | | 3 | 3 | | 9 |
| 3 | | | 3 | 3 | 3 | | | 9 |
| 4 | 3 | 6 | | | | | | 9 |
| 5 | | | 3 | 3 | 3 | | 6 | 15 |
| **Total** | 3 | 6 | 12 | 9 | 12 | 3 | 6 | 51 |

Supplementary material 2. Image of the device used in the fieldwork. A stainless steel structure equipped with two stereoscopic action cameras in horizontal position.

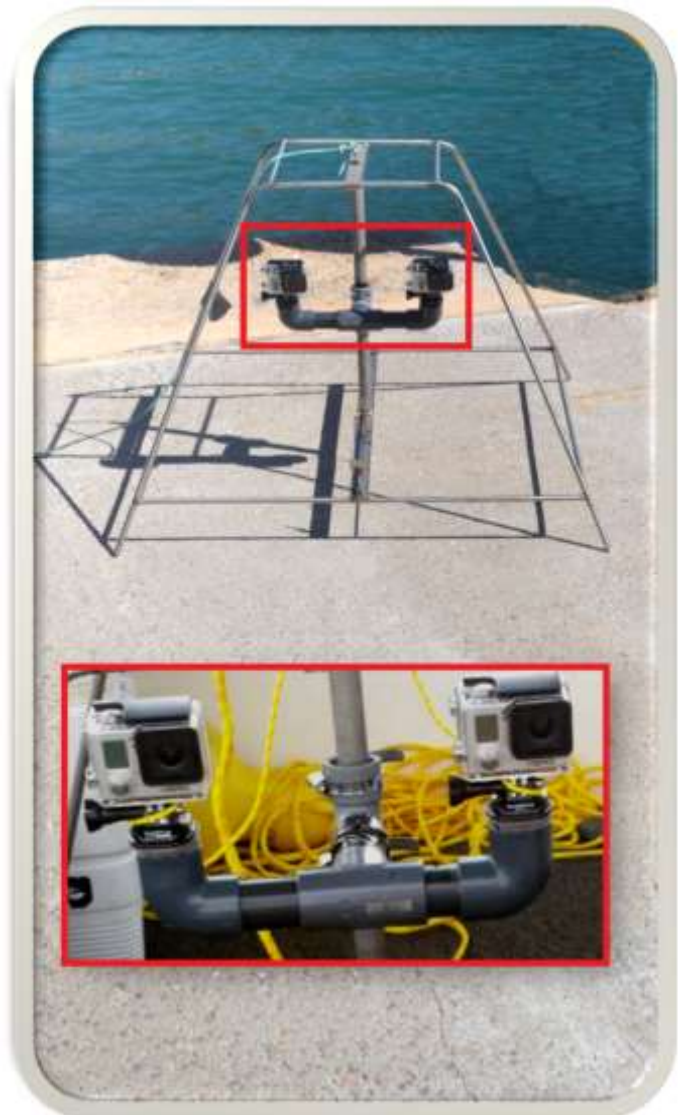

Supplementary material 3. Detail image of each point sampled with UVC and RUV. The number on the top-right matches the number in Table 1 of results.

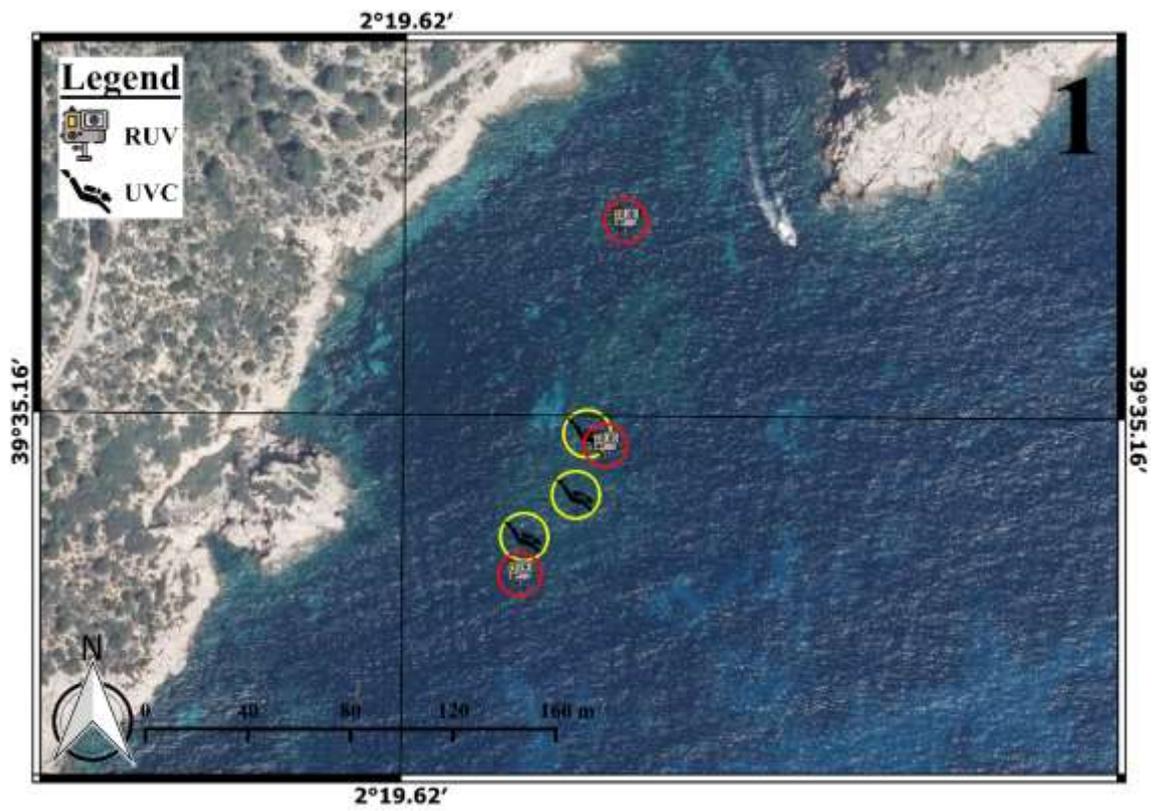

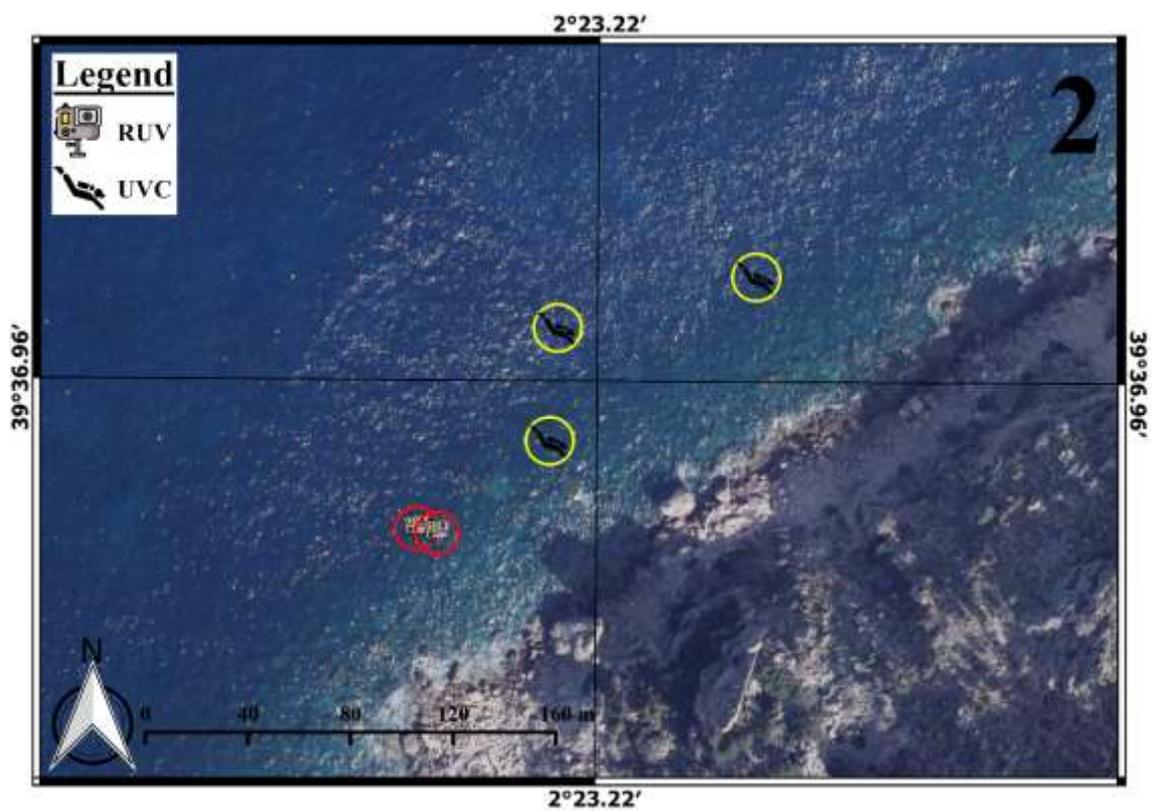

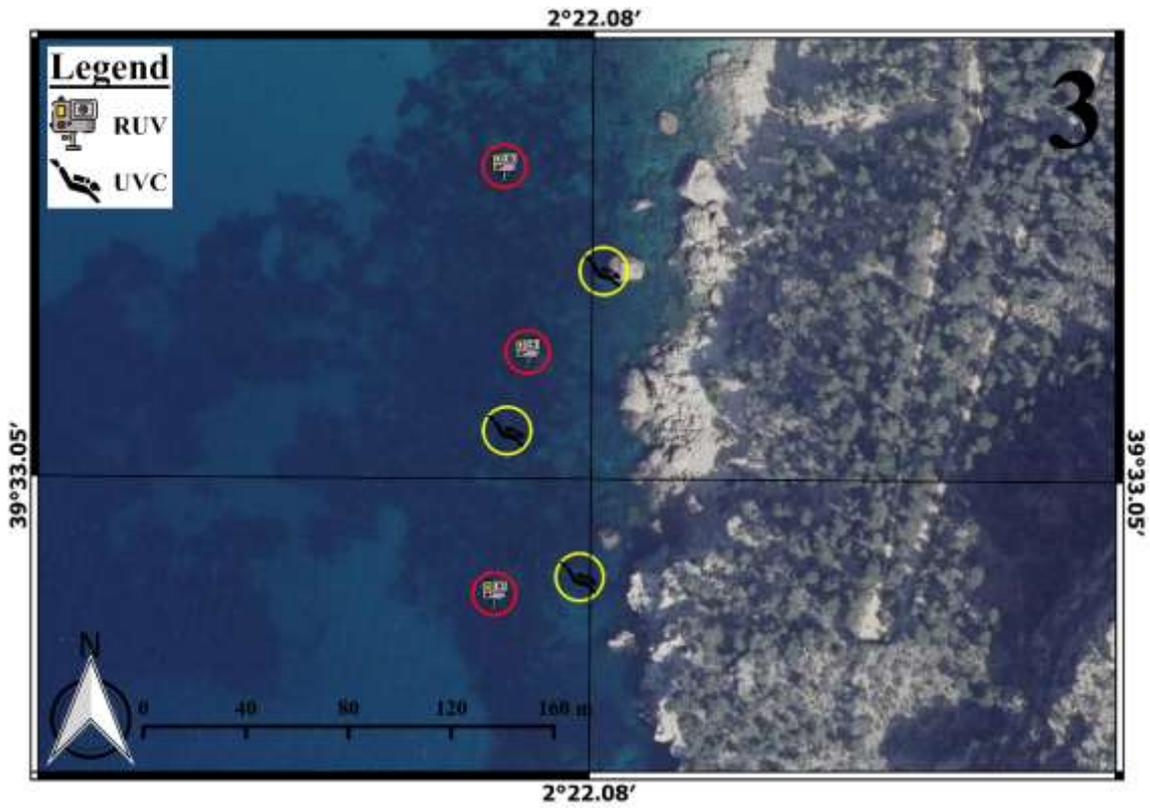

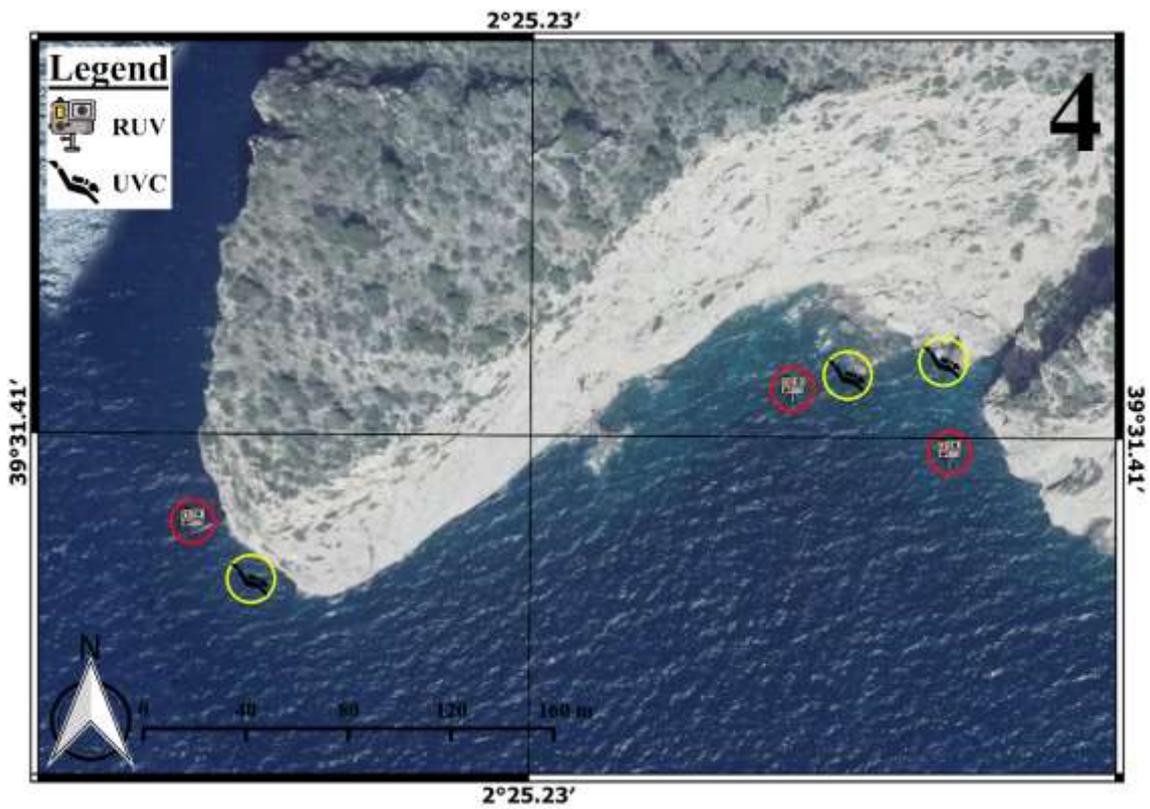

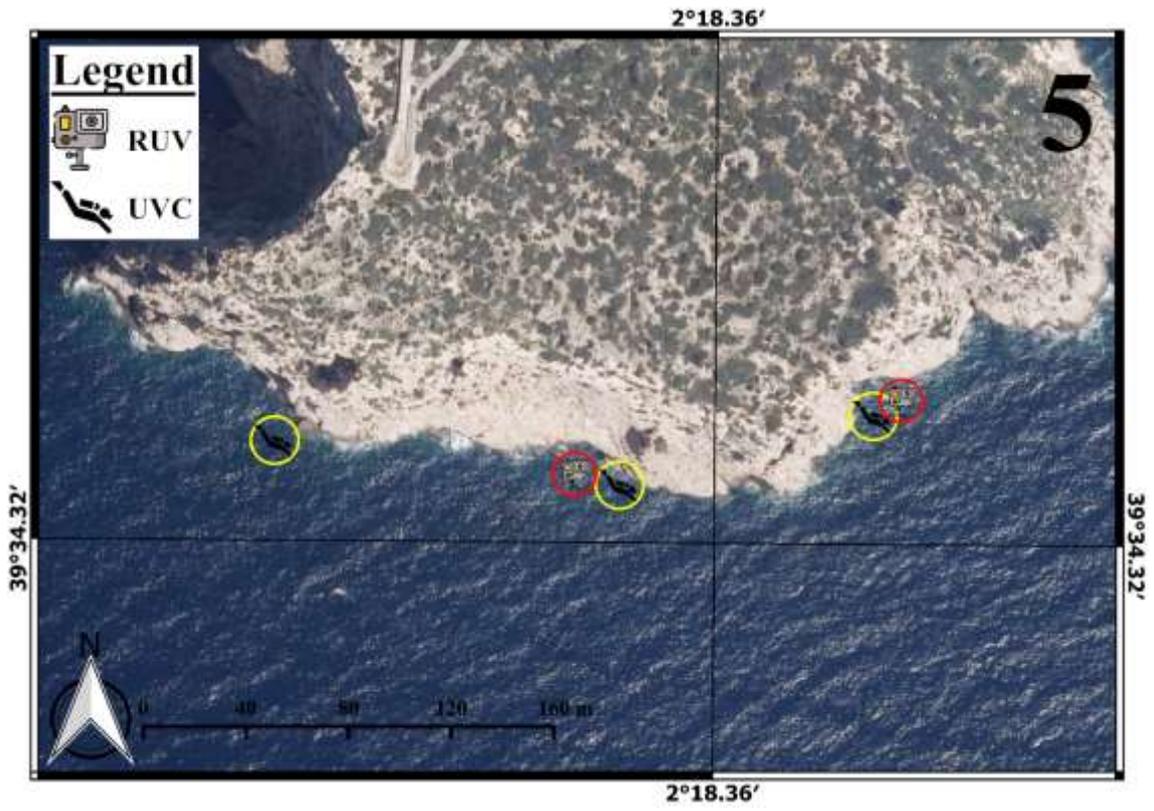

Supplementary material 4. Each letter represents the simulation of 5 virtual scenarios with a fixed density (1, 2, 5, 10, 20 ind 100m$^{-2}$) and a different individual detectability (boxplot) whit uncertainty. The simulation was 10 *virtual cameras* (i.e., a new *set* of camera-animals were created and moved to avoid interferences between cameras and improve computer sources) *replicated* 100 times (in the densities 10 and 20 ind 100m$^{-2}$ were 20 simulations *sets*) for estimating accuracy and precision. In each case, animal density was estimated at several moments (i.e., using an increasing number of frames, X-axis) by the Bayesian implementation of the model described above. The red line is the real density in the virtual scenario. Each green dot is the estimated density in the simulation of 10 cameras. The red dot is the mean of all the simulations in each moment. The accuracy is how much close the red point is to the red line and the precision is de dispersion of the green dots around the red dot.

Sample effort (frames*2,5min)

Sample effort (frames*2,5min)

Sample effort (frames*2,5min)

Sample effort (frames*2,5min)

E

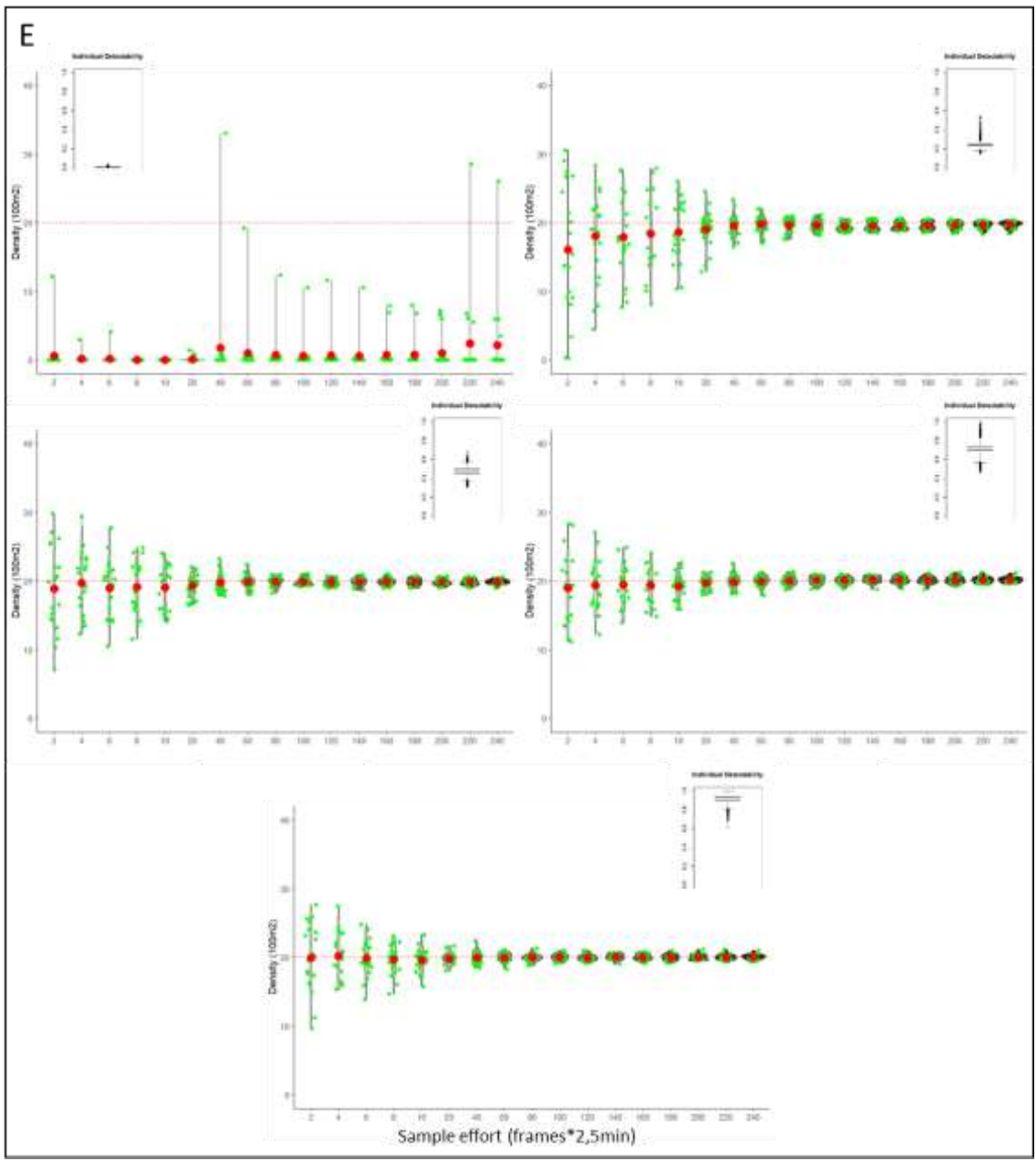

Sample effort (frames*2,5min)